\documentclass[oneside,twocolumn,aps,pra,preprintnumbers,bibnotes10pt,superscriptaddress,amsmath,amssymb]{revtex4-2}
\usepackage{graphicx,subfigure,wrapfig}
\usepackage{float}
\usepackage{color}
\usepackage{epstopdf}
\usepackage{amsmath}
\usepackage{braket}
\usepackage{amsthm}
\usepackage{amssymb}
\usepackage{amsfonts}
\usepackage{graphicx,subfigure}
\usepackage{txfonts}
\usepackage{ulem}
\usepackage{mathtools}
\usepackage{bm}
\usepackage{hyperref}
\usepackage{natbib}
\usepackage{comment}

\newcommand{\mean}[1]{\ensuremath{\big\langle #1 \big\rangle}}

\newcommand{\be}{\begin{equation}}
\newcommand{\ee}{\end{equation}}
\newcommand{\beq}{\begin{eqnarray}}
\newcommand{\eeq}{\end{eqnarray}}
\newcommand{\vect}[1]{\bm{#1}}

\begin{document}

\title{Stabilizing persistent currents in an atomtronic Josephson junction necklace} 

\author{L. Pezzè}
\altaffiliation{These authors contributed equally to this work}
\affiliation{Istituto Nazionale di Ottica del Consiglio Nazionale delle Ricerche (CNR-INO), Largo Enrico Fermi 6, 50125 Firenze, Italy}
\affiliation{European Laboratory for Nonlinear Spectroscopy (LENS), Via N. Carrara 1, 50019 Sesto Fiorentino, Italy}
\affiliation{QSTAR, Largo Enrico Fermi 2, 50125 Firenze, Italy}

\author{K. Xhani}
\altaffiliation{These authors contributed equally to this work}
\affiliation{Istituto Nazionale di Ottica del Consiglio Nazionale delle Ricerche (CNR-INO), Largo Enrico Fermi 6, 50125 Firenze, Italy}
\affiliation{European Laboratory for Nonlinear Spectroscopy (LENS), Via N. Carrara 1, 50019 Sesto Fiorentino, Italy}
\affiliation{QSTAR, Largo Enrico Fermi 2, 50125 Firenze, Italy}

\author{C. Daix}
\altaffiliation{These authors contributed equally to this work}
\affiliation{European Laboratory for Nonlinear Spectroscopy (LENS), Via N. Carrara 1, 50019 Sesto Fiorentino, Italy}
\affiliation{University of Florence, Physics Department, Via Sansone 1, 50019 Sesto Fiorentino, Italy}

\author{N. Grani}
\affiliation{Istituto Nazionale di Ottica del Consiglio Nazionale delle Ricerche (CNR-INO), Largo Enrico Fermi 6, 50125 Firenze, Italy}
\affiliation{European Laboratory for Nonlinear Spectroscopy (LENS), Via N. Carrara 1, 50019 Sesto Fiorentino, Italy}
\affiliation{University of Florence, Physics Department, Via Sansone 1, 50019 Sesto Fiorentino, Italy}

\author{B. Donelli}
\affiliation{Istituto Nazionale di Ottica del Consiglio Nazionale delle Ricerche (CNR-INO), Largo Enrico Fermi 6, 50125 Firenze, Italy}
\affiliation{University of Naples “Federico II”, Via Cinthia 21, 80126 Napoli, Italy}
\affiliation{QSTAR, Largo Enrico Fermi 2, 50125 Firenze, Italy}

\author{F.~Scazza}
\affiliation{University of Trieste, Physics Department, Via A. Valerio 2, 34127 Trieste, Italy}
\affiliation{Istituto Nazionale di Ottica del Consiglio Nazionale delle Ricerche (CNR-INO), Largo Enrico Fermi 6, 50125 Firenze, Italy}
\affiliation{European Laboratory for Nonlinear Spectroscopy (LENS), Via N. Carrara 1, 50019 Sesto Fiorentino, Italy}

\author{D. Hernandez-Rajkov}
\affiliation{Istituto Nazionale di Ottica del Consiglio Nazionale delle Ricerche (CNR-INO), Largo Enrico Fermi 6, 50125 Firenze, Italy}
\affiliation{European Laboratory for Nonlinear Spectroscopy (LENS), Via N. Carrara 1, 50019 Sesto Fiorentino, Italy}

\author{W.~J. Kwon}
\affiliation{Department of Physics, Ulsan National Institute of Science and Technology (UNIST), Ulsan 44919, Republic of Korea}

\author{G. Del Pace}
\affiliation{European Laboratory for Nonlinear Spectroscopy (LENS), Via N. Carrara 1, 50019 Sesto Fiorentino, Italy}
\affiliation{University of Florence, Physics Department, Via Sansone 1, 50019 Sesto Fiorentino, Italy}

\author{G. Roati}
\affiliation{Istituto Nazionale di Ottica del Consiglio Nazionale delle Ricerche (CNR-INO), Largo Enrico Fermi 6, 50125 Firenze, Italy}
\affiliation{European Laboratory for Nonlinear Spectroscopy (LENS), Via N. Carrara 1, 50019 Sesto Fiorentino, Italy}

\begin{abstract}

Arrays of Josephson junctions are at the forefront of research on quantum circuitry for quantum computing, simulation and metrology.
They provide a testing bed for exploring a variety of fundamental physical effects where macroscopic phase coherence, nonlinearities and dissipative  mechanisms compete. 
Here we realize finite-circulation states in an atomtronic Josephson junction necklace, consisting of a tunable array of tunneling links in a ring-shaped superfluid.
We study the stability diagram of the atomic flow by tuning both the circulation and the number of junctions. 
We predict theoretically and
demonstrate experimentally that the atomic circuit withstands higher circulations (corresponding to higher critical currents) by increasing the number of Josephson links. 
The increased stability contrasts with the trend of the superfluid fraction -- quantified by Leggett's criterion -- which instead decreases with the number of junctions and the corresponding density depletion.
Our results demonstrate atomic superfluids in mesoscopic structured ring potentials as excellent candidates for atomtronics applications, with prospects towards the observation of non-trivial macroscopic superpositions of current states. 

\end{abstract}

\maketitle

Josephson junction arrays are pivotal and versatile elements that hold promise to turn quantum mechanics into emerging computing, sensing and simulation technologies~\cite{YouPHYSTOD2005, ClarkeNATURE2008, DevoretSCIENCE2013, KjaergaardARCM2020, RasmussenPRXQuantum2021}. 
By harnessing the dissipationless non-linearity of single Josephson junctions, combined with strong collective effects, they show intriguing synchronization~\cite{JainPHYSREP1984, WiesenfeldPRE1998, VinokurNATURE2008, GrebenchukPRAPP2022} and interference~\cite{AndersonSCIENCE1998, CataliottiSCIENCE2001, IoffeNATURE2002} phenomena.
Furthermore, they serve as experimental tools to investigate the phase coherence and order parameters in high-$T_c$ superconductors~\cite{HilgenkampRMP2002, TafuriRPP2005}.

An array of junctions in a multiply-connected geometry forms a Josephson junction necklace~(JJN). 
In this configuration, the Josephson effect is used to control the current of persistent states, implementing robust dynamical regimes characterized by the competition between tunneling and interaction energies~\cite{BaroneBOOK}. 
JJNs with one or two junctions realize common quantum interference devices (SQUIDs)~\cite{ClarkeBOOK, FagalyRSI2006}, which find applications in rotation sensing with superfluid gyroscopes~\cite{SchwabNATURE1997, SatoRPP2012} and magnetic field sensing with superconducting rings~\cite{JaklevicPRL1964, ClarkeBOOK}. 
Furthermore, JJNs are key elements of atomtronic circuits~\cite{AmicoPRL2005, SeamanPRA2007, AmicoAVS2021, AmicoRMP2022}. 
Ultracold atoms in toroidal traps with a single junction or a weak link have been explored for the experimental realization of ultra-stable circulation states~\cite{RamanathanPRL2011, MoulderPRA2012, CaiPRL2022, DelPacePRX2022}, including the study of various superfluid decay phenomena~\cite{WrightPRL2013, WrightPRA2013, PoloPRL2019}, current-phase relations~\cite{EckelPRX2014} and hysteresis~\cite{EckelNATURE2014}. 
These experiments have stimulated several theoretical studies that have mainly focused on the analysis of different instability phenomena in ring superfluids with various types of defects and potentials~\cite{ WatanabePRA2009, PiazzaPRA2009, PiazzaPRA2010, DubessyPRA2012, MehdiSCIPOST2021,MateoPRA2015, PerezObiolPRE2020, xhani2023decay}.
In addition, double-junction atomtronic SQUIDs have enabled the observation of different regimes of Josephson dynamics~\cite{RyuPRL2013}, resistive flow~\cite{JendrzejewskiPRL2014} and quantum interference of currents~\cite{RyuNATCOM2020}.
Interestingly, as conjectured by Feynman~\cite{FeynmanBOOK}, further intriguing quantum coherence effects can arise -- due to the stiffness of the superfluid phase -- in ring systems hosting arrays of multiple junctions. 
However, despite advancements both in manufacturing mesoscopic nanostructured multi-linkcircuits~\cite{MeltzerSST2016, UriNANO2016, WolterSST2022, ZagoskinLTP2017, KornevLTP2017} and in engineering atomic trapping potentials~\cite{Gauthier2016,Zupancic2016,AmicoAVS2021,NavonNATPHYS2021}, the realization of tunable JJNs with arbitrary number of junctions remains technologically and experimentally challenging, and so far elusive in both superconducting and superfluid platforms.

In this work, we investigate supercurrent states in an atomtronic JJN. 
We analytically predict the stabilization of persistent currents against decay by increasing the number of junctions, $n$. 
We support this surprising prediction by numerical simulations and we demonstrate it experimentally in a bosonic superfluid ring with $n$ up to $16$.
Such an effect is a direct consequence of the single-valuedness of the order parameter, reflecting the  macroscopic phase coherence of the superfluid state.
Increasing the number of Josephson links leads to a decrease of the superfluid speed across each junction and to the corresponding increase of the global maximum (critical) current in the ring. 
Furthermore, the density depletion associated to an increasing $n$ determines a decrease of the superfluid fraction according to Leggett's formulation~\cite{LeggettPRL1970, LeggettJSP1998} that, however, does not result in a decrease of the critical current.
The full control of our atomtronic circuit opens exciting prospects toward the realization of non-trivial quantum superpositions of persistent currents~\cite{NunnenkampPRA2008, SolenovPRL2010, SchenkePRA2011, HallwoodPRA2011, AmicoSCIREP2014}.

\begin{figure}[b!]
\includegraphics[width=0.47\textwidth]{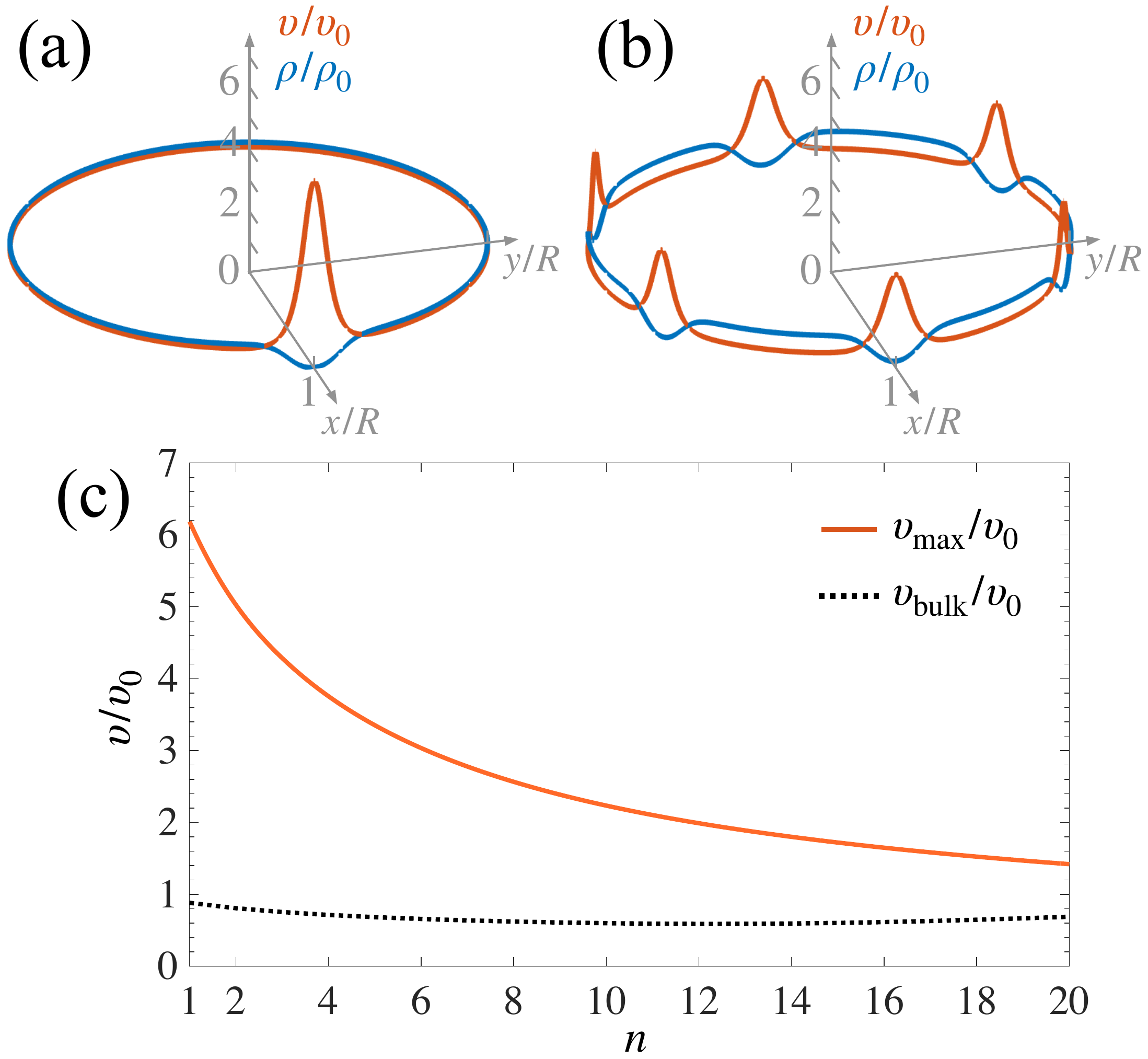}
\caption{{\bf Superfluid speed in a JJN}.
Panels~(a) and~(b) show the particle density $\rho$ (blue line) and the superfluid speed $\upsilon$ (orange line) in a 1D ring, divided
by the density ($\rho_0$) and speed ($\upsilon_0$) in the homogeneous ring, respectively. 
The two panels correspond to a one-dimensional JJN with $n=1$ (a) and $n=6$ (b) junctions, respectively. 
(c) Maximum, $\upsilon_{\rm max}$ (solid orange line), and bulk, $\upsilon_{\rm bulk}$ (dotted black line), superfluid speed as a function of the number of junctions.
Results in all panels are obtained from the stationary state of the one-dimensional GPE with $w=1$ and $\Omega=0$.
}
\label{Figure1}
\end{figure}

\section{Results}
 
{\bf Critical current in a multi-junction Josephson necklace.} 
A steady superfluid state can be described by a collective wavefunction $\psi(\vect{r}) = \sqrt{\rho(\vect{r})}e^{i \phi(\vect{r})}$, with $\rho(\vect{r})$ and $\phi(\vect{r})$ being the density and  the phase of the superfluid, respectively~\cite{LeggettRMP1999}. 
The latter is related to the superfluid speed by $\vect{\upsilon}(\vect{r}) = \tfrac{\hbar}{m} \nabla \phi(\vect{r})$, where $m$ is the atomic mass and $\hbar$ the reduced Planck constant.
To ensure a single-valued wavefunction, the integral of $\nabla \phi(\vect{r})$ calculated around any arbitrary closed path $\Gamma$ must be a multiple of $2\pi$,
\be \label{quantization}
\frac{m}{\hbar} \oint_\Gamma d \vect{r} \cdot  \vect{\upsilon}(\vect{r}) = 2 \pi w,
\ee
where the integer (winding) number $w$ is a topological invariant.
In a multiply-connected geometry (e.g. in a toroidal superfluid), Eq.~(\ref{quantization}) defines a series of quantized persistent-current states labeled by $w$~\cite{FetterPR1967, BlochPRA1973}.  
Although the ground state is $w=0$, metastable finite-circulation states ($w \neq 0$) can be generated, as first demonstrated with liquid helium~\cite{ClowPRL1967,ReppyJLTP1992} and more recently with ultracold atomic gases~\cite{RyuPRL2007, RamanathanPRL2011, BeattiePRL2013, MoulderPRA2012, KumarPRA2018, DelPacePRX2022}.

\begin{figure*}[t!]
\includegraphics[width=1\textwidth]{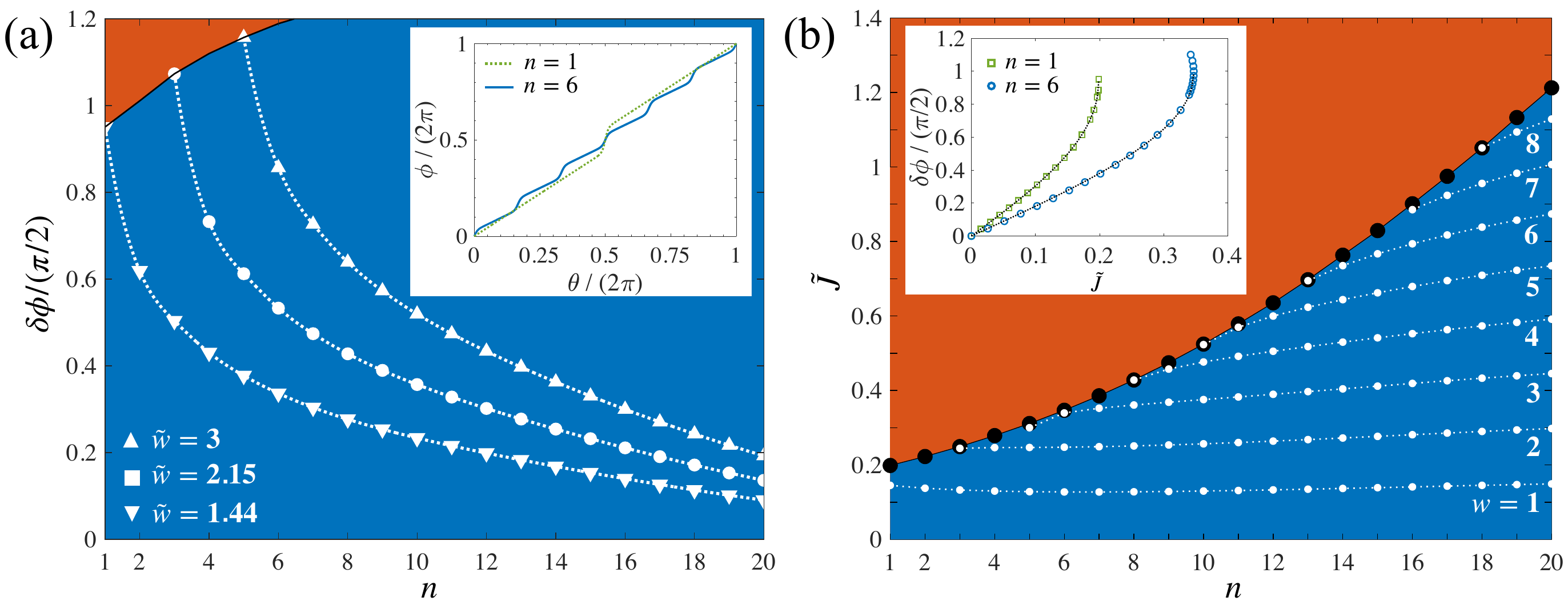}
\caption{{\bf Superfluid phase and critical current in a JJN}.
(a) Phase gain across each junction as a function of $n$, Eq.~(\ref{Eq.deltaphi}), where $f(\tilde{w},n)$ and $\rho_{\rm bulk}$ are obtained with GPE calculations.
Symbols are obtained for $\tilde{w}=1.44$ (downward triangles), $\tilde{w}=2.15$ (squares) and $\tilde{w}=3$ (upward triangles), which correspond to the maximum value of $\tilde{w}$ for $n=1$, 3 and 5, respectively, for which a stable solution can be found.
For larger values of $\tilde{w}$, for the given $n$, the system is unstable due to the nucleation of solitons.  
Lines are guides to the eye.
In particular, the solid black line connects maxima of $\delta \phi$ obtained for different $w$, separating the stable (blue) from the unstable (orange) region.
The inset shows the superfluid phase $\phi$ as a function of the angle $\theta$ along the ring, for $n=1$ (dotted green line) and $n=6$ (solid blue line).
(b) Critical current as a function of the number $n$ of junctions. 
The analytic formula Eq.~(\ref{Eq.Jc})~(large black dots) superpose to the numerical calculation of the maximum current.
Small white dots show the current $\tilde{J}$ calculated for $\Omega=0$ and different values of $w$, ranging  from $w=1$ (lower) to $w=8$ (upper).
Solid and dotted lines are guides to the eye.
The orange region corresponds to values of the current above $\tilde{J}_c$ and are thus inaccessible in the system. 
Inset: phase across each junction as a function of the current (symbols) for $n=1$ (green squares) and $n=6$ (blue circles).
The solid lines are the current-phase relations $\delta \phi = \arcsin(\tilde{J}/\tilde{J}_c) - 2\pi\ell \tilde{J}$, with $\tilde{J}_c$ and $\ell$ extracted from fitting.} 
\label{Figure2}
\end{figure*}

Let us consider, for the sake of illustration, a one-dimensional (1D) JJN of radius $R$ with $n$ equivalent junctions modelled as narrow Gaussian potential barriers, rotating with angular velocity $\Omega$ (see \cite{supp}).
In the rotating frame, the current of stationary states is given by 
\be \label{Jgeneral}
J = \rho(\theta) [\upsilon(\theta)/R - \Omega],
\ee
where $\theta$ is the azimuthal angle along the ring. 
Each junction induces a dip in the particle density $\rho(\theta)$, shown as the blue line in Fig.~\ref{Figure1}(a) and (b) as calculated from the stationary state of the one-dimensional Gross-Pitaevskii equation (GPE \cite{supp}).
We emphasize that the barrier height is larger than the chemical potential and the barriers width is of the order of the superfluid healing length~\cite{supp}, with the density not vanishing inside the barrier.
Due to the conservation of mass-flow [see Eq.~(\ref{Jgeneral})] a density dip implies a local increase of the superfluid speed $\upsilon(\theta)$ [orange lines in Fig.~\ref{Figure1}(a) and (b)].
Comparing the panels (a) and (b) of Fig.~\ref{Figure1}, obtained for the same value of the circulation $w$ and for different number of junctions, $n=1$ and $n=6$, respectively, we observe that the maximum superfluid speed, $\upsilon_{\rm max}$, drops by increasing $n$.
This is a consequence of the topological invariance expressed by Eq.~(\ref{quantization}). 
This is seen by writing $\upsilon(\theta) = \upsilon_{\rm bulk} + \upsilon_{n-{\rm peaks}}(\theta)$, where $\upsilon_{\rm bulk}$ is the bulk speed, given by the minimum velocity along the ring, and $\upsilon_{n-{\rm peaks}}(\theta)$ describes the $n$ peaks of the superfluid speed. 
Replacing this expression for $\upsilon(\theta)$ into Eq.~(\ref{quantization}), we find
\be \label{Eq.speed}
 \upsilon_{\rm bulk} + \frac{1}{2\pi} \int_{0}^{2\pi} d\theta\ \upsilon_{n-{\rm peaks}}(\theta) = \frac{\hbar w}{mR}.
\ee
The bulk contribution in Eq.~(\ref{Eq.speed}) is expected to change only slightly when adding sufficiently-narrow junctions to the JJN [see the dotted black line in Fig.~\ref{Figure1}(c)]. 
On the contrary, the second term in Eq.~(\ref{Eq.speed}) is proportional to $n\upsilon_{\rm max}$.
Therefore, for a given $w$, $\upsilon_{\rm max}$ must decrease roughly as $1/n$ in order to keep the integral in Eq.~(\ref{Eq.speed}) constant. 
The decrease of $\upsilon_{\rm max}$ is confirmed by the results of GPE simulations reported in Fig.~\ref{Figure1}(c) [solid orange line].
The reduction of the superfluid velocity at each barrier implies a decrease of the phase gain $\delta \phi$ across each junction, upon increasing $n$.
For a more quantitative study, we use Eq.~(\ref{Eq.speed}) and notice that $\upsilon_{\rm bulk} = JR/\rho_{\rm bulk} + \Omega R$, where $\rho_{\rm bulk}$ is the bulk density, given by the maximum density along the ring, and $n\delta \phi = (mR/\hbar) \int d\theta \, \upsilon_{n-{\rm peaks}}(\theta)$.
We find the relation
\be \label{quantizationJ}
\frac{2 \pi \tilde{J}}{\rho_{\rm bulk}} + n \delta \phi = 2 \pi \tilde{w}, 
\ee
where $\tilde{J}=J/J_R$, $J_R = \hbar/(mR^2)$ is the current quantum in the homogeneous (no junctions) case and $\tilde{w} = w - \Omega/J_R$ is an effective circulation in the rotating frame.
Varying $\Omega$ allows to address non-integer $\tilde{w}$ and thus continuous values of the current.
Furthermore, by inserting Eq.~(\ref{Jgeneral}) into Eq.~(\ref{quantization}), we obtain 
\be \label{Eq.J}
\tilde{J}= \frac{\tilde{w} f(\tilde{w},n)}{2\pi},
\ee
where $f(\tilde{w},n) = (2\pi)^2 \big[\int d\theta/\rho(\theta)\big]^{-1}$.
We note that $f(\tilde{w},n) \leq f_s$, where $f_s \in [0, 1]$ is Leggett's superfluid fraction~\cite{LeggettPRL1970, LeggettJSP1998, ZapataPRA1998, ChauveauPRL2023, TaoPRL2023, Biagioni}.
The latter expresses the phase rigidity of the system, quantified by the kinetic-energy response to a phase twist of the superfluid order parameter.
Since $f(\tilde{w},n)=f_s$ for $w=0$ and in the limit $\Omega \to 0$~\cite{supp}, Eq.~(\ref{Eq.J}) connects the superfluid fraction to the current in the ring.
It is possible to see that $f_s$ decreases with $n$ as far as the junctions do not overlap substantially~\cite{supp}, therefore, according to Eq.~(\ref{Eq.J}) the current decreases as well. 

On the other hand, by combining Eqs.~(\ref{quantizationJ}) and (\ref{Eq.J}), 
the phase across each junction reads
\be \label{Eq.deltaphi}
\delta \phi = \frac{2\pi \tilde{w}}{n} \bigg( 1 - \frac{f(\tilde{w},n)}{\rho_{\rm bulk}} \bigg).
\ee
In Fig.~\ref{Figure2}(a), we plot $\delta \phi$ as a function of $n$, Eq.~(\ref{Eq.deltaphi}), where the quantities $f(\tilde{w},n)$ and $\rho_{\rm bulk}$ are obtained from the stationary states of the 1D GPE. 
Symbols refer to different values of $\tilde{w}$.
We clearly see that $\delta \phi$ decreases with $n$.
This implies that the condition $\delta\phi \approx \pi/2$~\cite{BaroneBOOK, TilleyBOOK} -- that determines the maximum (or critical) current $\tilde{J}_c$ in the JJN -- for increasing $n$ is met for higher values of $\tilde{w}$. 
We find an explicit expression for $\tilde{J}_c$, by considering the usual current-phase relation $\delta \phi = \sin^{-1} (\tilde{J}/\tilde{J}_c) - 2 \pi \ell \tilde{J}$ \cite{BaroneBOOK, WrightPRL2013}, with $\ell$ an adimensional kinetic inductance. 
The condition $\tilde{J}= \tilde{J} _c$  provides a critical phase $\delta \phi_c = \pi/2 - 2 \pi \ell \tilde{J}_c$.
Replacing this value into Eq. (\ref{Eq.deltaphi}) and using Eq.~(\ref{Eq.J}), we find 
\be \label{Eq.Jc}
\tilde{J}_c = 
\frac{n f_c/4}{2\pi (1 - f_{c}/ \rho_c ) + n f_c \ell},
\ee
where $f_{c}$ and $\rho_c$ are the values of $f(\tilde{w},n)$ and $\rho_{\rm bulk}$ obtained when $\tilde{J}=\tilde{J}_c$.
Neglecting the small inductance ($n \ell \ll 2\pi$), we find that the critical current is mainly determined by the competition between $n$ and $f_c(n)$.
In Figure~\ref{Figure2}(b) we plot the critical current obtained from the GPE solution as a function of $n$. 
Numerical values agree with Eq.~(\ref{Eq.Jc})
(black dots, with the solid line being a guide to the eye), where $\ell$ is extracted from a fit of the numerical current-phase relation, e.g. shown in the inset for $n=1$ (green squares) and $n=6$ (blue circles).
Furthermore, small white dots in Fig.~\ref{Figure2}(b) show the current of metastable states in the case $\Omega = 0$, where $\tilde{J}$ assumes only quantized values, see Eq.~(\ref{Eq.J}) with $\tilde{w}=w$.
Figure.~\ref{Figure2}(b) clearly shows that $\tilde{J}_c$ increases with the number of junctions. 
When $\tilde{J} > \tilde{J}_c$, the current enters the unstable regime [red regions in Fig.~\ref{Figure2}(a)-(b)], characterized by the simultaneous emission of $n$ solitons from the barriers (see Refs.~\cite{MateoPRA2015, PerezObiolPRE2020} for a study of the case $n=1$).

\begin{figure*}[t!]
\includegraphics[width=1\textwidth]{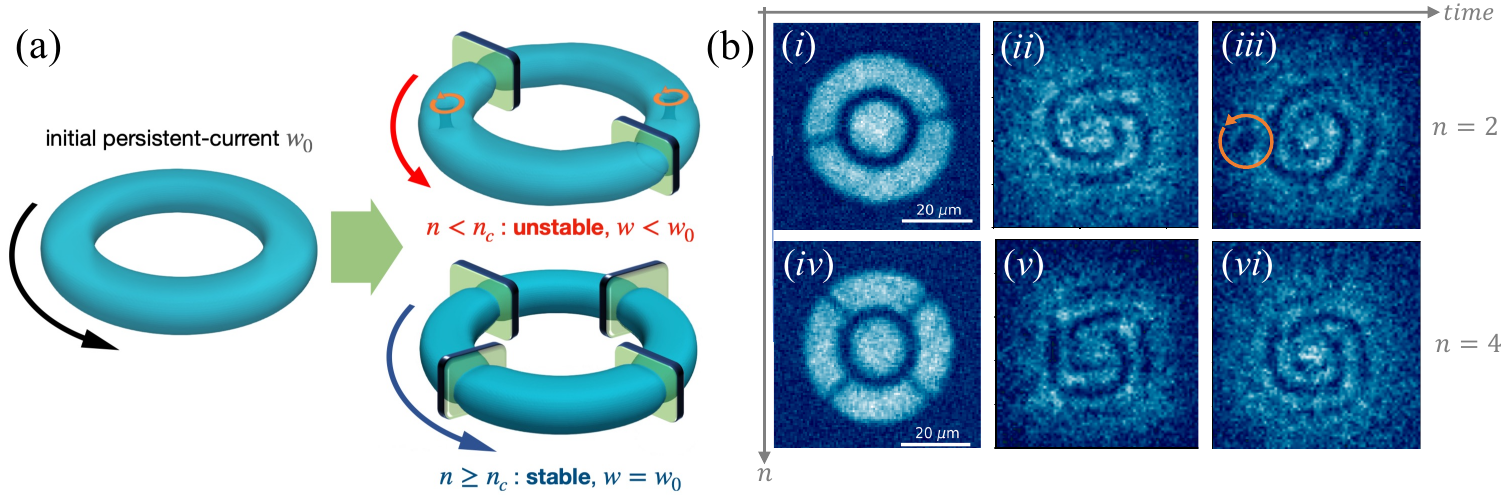}\caption{{\bf Sketch of the experiment and observables.} 
(a) After preparing an initial persistent current state with circulation $w_0$, the $n$ junctions are ramped up (see text).
The 3D density plots are isosurfaces obtained from 3D GPE numerical simulations of the experimental set-up.
If $n$ is below a critical value $n_c$ depending on $w_0$, the initial current is dissipated via the nucleation of vortices (here $n=2$ and vortices are highlighted by orange cycling arrows in the upper right plot). 
Conversely, if $n\geq n_c$ (here $n=4$), the system remains stable with $w = w_0$ (lower right plot). 
(b) Examples of single-shot experimental in-situ images and interferograms obtained for $w_0=2$ and for the same number of junctions $n$ as in (a):
$n=2$ (unstable configuration), at $t=0$ ({\it i}), $t=1$ ms ({\it ii}) and $t=7$ ms ({\it iii}); and
$n=4$ (stable configuration) for $t=0$ ms ({\it iv}), $t=1$ ms ({\it v}) and $t=20$ ms ({\it vi}).
In the case ({\it iii}), the circulation has decayed ($w(t)<w_0$) and the vortex emission is identified by the single spiral arm and the presence of a localized region of low density, i.e. a vortex.
}
\label{Figure3}
\end{figure*}

Although the above discussion is restricted to a 1D geometry, the predicted effects are expected to hold also in higher dimensions, due to the general validity of Eq.~(\ref{quantization}).
To confirm this expectation, we have performed 3D time-dependent Gross-Pitaevskii simulations~\cite{supp}. 
We prepare the ground state in an annular trap, impose a circulation $w_0$, and observe the dynamics of the system in the presence of $n$ junctions.
Consistently with the results of Fig.~\ref{Figure2}, we observe a decrease with $n$ of both the superfluid speed and the time-averaged phase gain across each junction~\cite{supp}. 
The results of numerical simulations are schematically summarized as in Fig.~\ref{Figure3}(a). 
If the number of junctions is below a critical value $n_c$ that depends on $w_0$, then vortices are emitted symmetrically from each barrier, causing phase slippage and a decay of both the current and the winding number in time~\cite{supp}. 
This vortex emission is the 3D analogue of the observed simultaneous nucleation of $n$ solitons in 1D simulations in the unstable regime.
If $n$ is increased above $n_c$, then the emission of vortices is suppressed and the circulation remains constant in time.
A higher stable circulation corresponds to a larger critical current. 

{\bf Experimental system and persistent current states.} 
We investigate experimentally the predicted increase of current stability in JJNs by realizing a Bose-Einstein condensate (BEC) of $^6$Li molecules in an annular trap equipped with a variable number ($n \leq 16$) of static planar junctions.
Both the ring-shaped trap and the array of junctions are produced by the same digital micromirror device (DMD) illuminated with blue-detuned light to provide a repulsive optical potential.
Using the high resolution of the DMD projection setup, we create a dark ring-shaped region in the $x$-$y$ plane delimited by hard walls whose height is much larger than the chemical potential of the superfluid (given by $\mu/h \simeq 850$ Hz in the clean ring), with $R_{\rm in} \simeq 11.7 \pm 0.2$ $\mu$m and $R_{\rm r out} \simeq 20.6 \pm 0.2$ $\mu$m being the inner and outer radius of the annulus. 
The potential is completed by a tight harmonic confinement along the vertical $z$ direction, of trapping frequency $\omega_z = 2 \pi \times (383 \pm 2)$ Hz. 
The junctions can be modelled as Gaussian peaks of initial height $V_0 \simeq (1.3 \pm 0.2)\,\mu$ and $1/e^2$-width $\sigma = (1.2 \pm 0.2)\,\xi$, with $\xi \approx 0.68~\mu$m being the healing length (see Ref.~\cite{supp} for details on the barrier characterization).
We initially trap $\simeq 6.8 \times 10^3$ condensed atom pairs inside the ring with a shot-to-shot stability around $5\%$. 
Due to the finite lifetime of our molecular BEC, 
the pair number decreases over the course of the current decay by at most $20 \%$, causing a decrease of the chemical potential of the superfluid. 
Consequently the value of $V_0/\mu$ increases by up to $\sim 15\%$ depending on the holding time.

\begin{figure*}[t!]
    \centering
    \includegraphics[width=\textwidth]{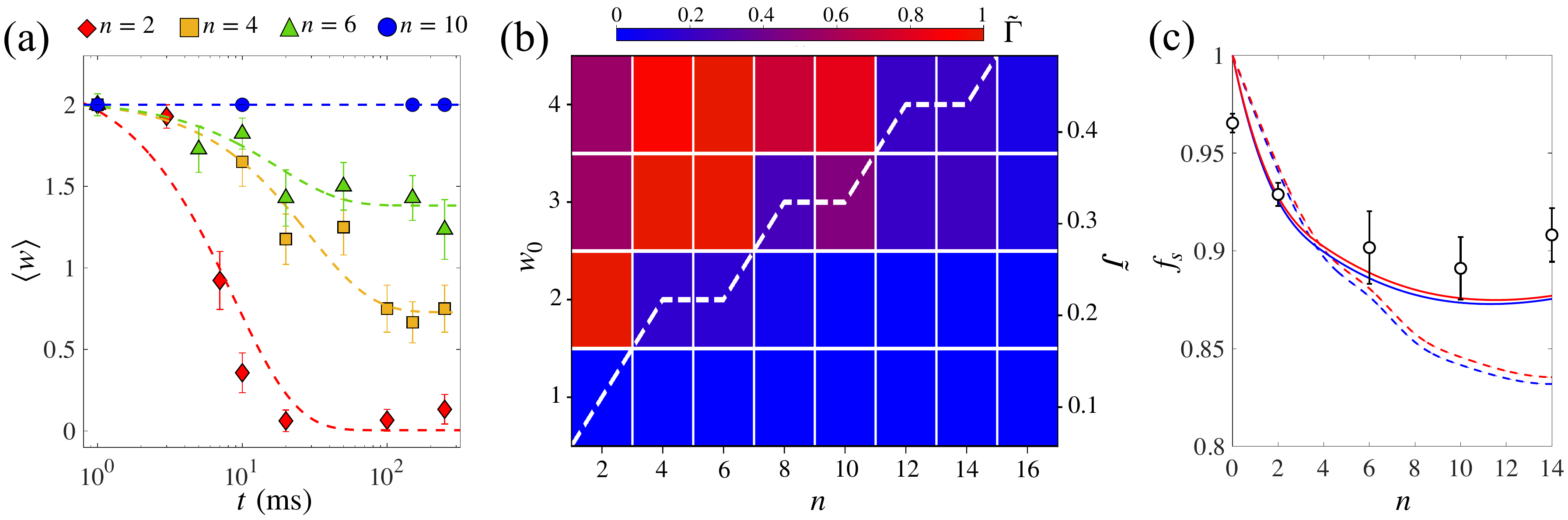}
    \caption{\textbf{Stability phase diagram of an atomtronic JJN.}
(a) Mean circulation as a function of time, for $w_0=2$ and different number of barriers, $n$ (symbols), with averages and error bars obtained from $\sim 15$ repeated measurements for each point.
The dashed lines are exponential fits, $\mean{w(t)}=w_f+\Delta w \exp{(-\Gamma t)}$.
(b) Effective decay rate $\tilde{\Gamma} \propto \Delta w \,\Gamma$ (colormap), extracted from the exponential fits as in panel (a) as a function of $w_0$ and $n$. 
$\tilde{\Gamma}$ quantifies the stability of an initial finite-circulation state $w_0$. 
The dashed white line is the critical circulation $w_{\mathrm c}(n)$ and the corresponding current (right axis) as a function of $n$, obtained from 3D GPE simulations.
(c) Upper (dashed red line) and lower (dashed blue line) bounds to the superfluid fraction $f_s$, Eq.~(\ref{fs_bounds}), as a function of the number of junctions.
Bounds are obtained from the ground state density of the numerical GPE. 
The solid lines are the bounds evaluated by including the finite resolution of the experimental imaging system.    
Circles are the upper bound evaluated using experimental in-situ images and averaged over 10 realizations.
}
\label{Figure4}
\end{figure*}

We initialize the superfluid ring in a quantized circulation state with winding number $w_0\in \{ 1,2,3,4\}$. 
Following the procedure described in Ref.~\cite{DelPacePRX2022}, different values of $w_0$ are obtained on-demand by shining a DMD-made azimuthal light intensity gradient onto the ring over a duration $t_I \ll \hbar/\mu$, i.e.~shorter than the characteristic density response time, $\hbar/\mu$. 
In this way, we imprint a phase $\Phi(\theta) = U_0(\theta)\times t_I/\hbar$ to the condensate wavefunction without modifying the atomic density~\cite{KumarPRA2018}, where $U_0(\theta)$ is the spin-independent potential exerted by the light field on the atomic states that varies linearly with $\theta$~\cite{DelPacePRX2022}.
After the imprinting, we wait 300 $\rm{ms}$ to let the cloud reach equilibrium, allowing the possible density excitations following the imprinting procedure to damp out \cite{xhani2023decay}.
We then progressively ramp up the $n$ Gaussian junctions over approximately 1\,ms (corresponding to $\approx 6 \, \hbar/\mu$).

{\bf Stability phase diagram.}
To measure the winding $w$ in the ring, we exploit an interferometric probe~\cite{EckelPRX2014, CormanPRL2014,DelPacePRX2022}: we equip the atomic superfluid with a central disk acting as a phase reference [see panels ({\it i}) and ({\it iv}) in Fig.~\ref{Figure3}(b)] and measure the relative phase between the disk and the ring from the interference pattern arising after a short time-of-flight.
The number of spiral arms in the interferogram provides access to the value of the circulation (winding number) at time $t$, $w(t)$.
The different panels of Fig.~\ref{Figure3}(b) display typical examples of experimental images.
In panels ({\it i}) and ({\it iv}) we show the in-situ atomic density profile at $t=0$.
The atomic density 
(averaged over 10 experimental images) is characterized by a homogeneous bulk both in the azimuthal and radial directions.
The $n=2$ ({\it i}) and $n=4$ ({\it iv}) junctions are clearly visible and are associated to local dips in the density, similarly as in Fig.~\ref{Figure1} and Fig.~\ref{Figure3}(a).  
In panels ({\it ii}) and ({\it iii}) we show examples of spiral interference patterns emerging for an unstable dynamics, namely $w(t)$ decreasing in time below $w_0$ (here, $w_0=2$ and $n=2$):
in ({\it ii}) $t=1$ ms and $w(t)=2$, while in ({\it iii}) $t=7$ ms and $w(t)=1$.
In particular, panel ({\it iii}) shows the presence of a vortex identified as a localized low-density defect and marked by the orange arrow.
The vortex emission signals the decrease of $w$ by one quantum.
In panels ({\it v}) and ({\it vi}) we show instead the interferograms for stable dynamics, namely $w(t)=w_0$ (here, $w_0=2$ and $n=4$).
A non-circular, polygonal interference pattern is visible both at short [({\it v}), $t=1$ ms] and at long [({\it vi}), $t=20$ ms] times due to the sharp phase gain at the junctions. 

By averaging the winding number over $\sim 15$ experimental realizations under the same conditions,  
we extract the evolution of the mean circulation $\mean{w(t)}$ for various $n$. 
We study the dynamics up to 250 ms, which is sufficient to observe steady current states at long-times while still limiting particle losses. 
The measured $\mean{w(t)}$ is shown in Fig.~\ref{Figure4}(a) for $w_0=2$.
We fit each curve with an exponential decay given by $\mean{w(t)}=w_f+\Delta w \exp{(-\Gamma t)}$. 
The fitting parameters $w_f$, $\Delta w$ and $\Gamma$ allow us to characterize the mean supercurrent. 
As $\mean{w(t)}$ is obtained from statistical averaging, the figure shows that the number of realizations $w(t)$ that remain stable in time increases with the number of junctions. 
In particular, the number of stable realizations increases substantially when changing the number of junctions from $n=2$ (red diamonds) to $n=4$ (yellow squares).
For $n=10$ (blue circles), all realizations 
are stable: this demonstrates the experimental capability to create stable finite-circulation states in a JJN.
  
Figure~\ref{Figure4}(b) summarizes the results obtained for different $w_0$ and $n$, in the form of a stability phase diagram.
In particular, we plot the quantity $\tilde{\Gamma} = \Delta w\,\Gamma/\max_n( \Delta w\,\Gamma)$, where each horizontal line of the phase diagram is normalized to its maximum value for fixed $w_0$.
This quantity combines information on the difference between the initial and the final winding numbers, $\Delta w$, namely how much the currents decay, and on the timescale over which this decay takes place, $\Gamma$. 
Values of $\tilde{\Gamma}\approx 1$ (red regions) are obtained when most of the realizations $w(t)$ rapidly decay towards values of the circulation lower than the initial $w_0$. 
On the contrary, small values of $\tilde{\Gamma}\approx 0$ (blue regions) are obtained when most of the realizations are stable over time, namely $w(t)=w_0$.
The phase diagram clearly shows that, on average, the system supports a higher number of stable realizations when increasing the number of junctions~\cite{supp}.
By the choice of normalization, $\tilde{\Gamma}$ shows a sharp transition from $\tilde{\Gamma}\approx 1$ to $\tilde{\Gamma}\approx 0$ when increasing $n$.
The dashed white line in Fig.~\ref{Figure3}(b) denotes the critical winding number $w_c(n)$ and the corresponding current (right axes) as a function of $n$,
as computed numerically from 3D GPE simulations.
The numerical critical curve $w_c(n)$ is obtained for $V_0/\mu = 1.8$ and match the experimental phase diagram well.
The need for a larger $V_0/\mu$ in numerical simulations with respect to the one estimated in the experiment, is consistent with the finite lifetime of the sample (which implies that $V_0/\mu$ increases during the dynamics) and the finite resolution of the DMD potential, which makes the barriers not perfectly identical~\cite{supp}.
Anyway, we note that the only effect of a change of $V_0/\mu$ on the critical line $w_c(n)$ is to provide a linear shift, meaning that the particular choice of $V_0/\mu$ does not affect its trend, which  
well reproduce the experimental findings. 

Given that $\tilde{J}_c(n) \sim n\, f_c(n)$ from Eq.~(\ref{Eq.Jc}), a significant decrease of the superfluid fraction $f_s \geq f_c$ would overshadow the stabilization mechanism arising from increasing $n$.  
For this reason, in Fig.~\ref{Figure4}(c), we study the dependence of $f_s$ on $n$ and indeed find a mildly decreasing trend,
which is insufficient to disrupt the enhanced stability of currents for large $n$.
According to a variational calculation by Leggett~\cite{LeggettPRL1970, LeggettJSP1998}, the superfluid fraction $f_s$ can be bounded experimentally from the in-situ density profile~\cite{ChauveauPRL2023, TaoPRL2023, Biagioni}:
\begin{equation}\label{fs_bounds}
    \iint\!\frac{dz\,dr~r}{\frac{1}{d^2}\int_{\rm cell} \frac{d\theta}{\rho(r,\theta,z)}} \leq f_s \leq \left( \frac{1}{d^2}\int_{\rm cell}\frac{d\theta}{\iint dz\, dr~r \rho(r,\theta,z)} \right)^{-1},
\end{equation}
where the density $\rho(r,\theta,z)$ is calculated from the ground state of the 3D GPE.
The bounds in Eq.~(\ref{fs_bounds}) are computed by restricting the azimuthal angle $\theta$ over a unit cell of size $d=2\pi/n$ and using the normalization $\iint dz\, dr~r \int_{\rm cell} d\theta \, \rho(r,\theta,z)=1$~\cite{LeggettPRL1970, LeggettJSP1998, ZapataPRA1998}.
In Fig.~\ref{Figure4} 
we plot the upper (dashed red line) and lower (dashed blue line) bounds in Eq.~(\ref{fs_bounds}).
They are very close to each other as our system is approximately separable in the transverse spatial directions~\cite{ChauveauPRL2023} and coincide in 1D, where $f_s = \lim_{w=0,~ \Omega \to 0} f(\tilde{w},n)$~\cite{supp}. 
Increasing $n$ enhances the size of the density dip relative to the unit cell length and thus decreases both the lower and upper limits in Eq.~(\ref{fs_bounds}), see Fig.~\ref{Figure4}(c). 
Experimentally, for each value of $n$, we compute Leggett's upper bound on 10 different images of the experimental density. 
We compute the integral on the right-hand side of Eq.~\eqref{fs_bounds} by summing over all pixels inside an annular region with inner and outer radii $r_{cut1}>R_{in}$ and $r_{cut2}<R_{out}$ respectively. 
We have numerically verified that the values of the bounds do not depend on the exact size of this region.
The corresponding mean values and standard deviations are shown as circles in Fig.~\ref{Figure4}(a).
The deviations from $f_s=1$ in the clean torus ($n=0$) are mainly due to noise in the experimental images, as well as the finite pixel size of our imaging sensor. 
Experimental results are well reproduced when taking into account the finite resolution of the imaging system (solid blue and red lines) and clearly show a decrease of $f_s$ with $n$.

\section{discussion}

Our work showcases the first experimental observation of ring supercurrents in periodic arrays of Josephson junctions. 
Such stable currents can be experimentally observed only for a sufficiently large number of links, as predicted by our theory modeling.    
In particular, our work shows that the maximum current flowing across the atomtronic circuit is due to a cooperative mechanism involving all the junctions rather than only to the properties of the single Josephson link.
We expect the mechanism demonstrated in this manuscript to apply to any superfluids and superconductors as it soleley depends on the single valuedness of the wavefunction in a multiply-connected topology.

Therefore, a natural extension of our work will be to investigate whether the same effect stabilizes supercurrents in other annular systems, such as atomic Fermi superfluids~\cite{CaiPRL2022, DelPacePRX2022} and supersolids~\cite{TengstrandPRA2021}.
In the former case, the condensate fraction differs from unity even at $T=0$~\cite{KwonSCIENCE2020} and additional dissipative effects, such as Cooper pair-breaking~\cite{WlazlowskiPRL2023, PisaniARXIV} may compete with the stabilization mechanism.
In the latter, intrinsic density modulations realize an array of self-induced Josephson junctions -- as recently demonstrated in Ref.~\cite{Biagioni} for an elongated atomic system -- which can be controlled by tuning the confinement parameters.

Finally, the exquisite controllability offered by our platform opens the way toward realizing exotic quantum superposition of superflow states~\cite{NunnenkampPRA2008, SolenovPRL2010, SchenkePRA2011, HallwoodPRA2011, AmicoSCIREP2014} with possible implications in both atomtronic and quantum technologies. \\

\noindent We thank Massimo Inguscio, Giovanni Modugno, Augusto Smerzi and Andrea Trombettoni for discussions. 
L.~P. and K.~X. acknowledge financial support by the QuantEra project SQUEIS. F.~S.~acknowledges funding from the European Research Council (ERC) under the European Union’s Horizon 2020 research and innovation programme (Grant agreement No.~949438) and from the Italian MUR under the FARE programme (project FastOrbit). G.~R.~acknowledges funding from the Italian Ministry of University and Research under the PRIN2017 project CEnTraL. G.~R. and G.~D.~P.~acknowledge financial support from the PNRR MUR project PE0000023-NQSTI. \\

\newpage

\section{Supplementary Information.}

{\bf Numerical methods.} We discuss here numerical methods used to obtain the results discussed in the main text. 

{\it 1D GPE.}
The 1D simulations shown in Fig.~\ref{Figure1} and \ref{Figure2} refer to static solutions of the GPE equation:
\be \label{GPEstationary}
\tilde{\mu} f(\theta) = \bigg( -\frac{1}{2} \frac{\partial^2}{\partial \theta^2} - \frac{1}{2} 
\frac{\tilde{J}^2}{\rho(\theta)^2} + \tilde{V}(\theta) + \tilde{g} \rho(\theta) \bigg) f(\theta),
\ee
where $f(\theta) = \sqrt{\rho(\theta)}$, $\tilde{J} = 2 \pi (w-\tilde{\Omega})/\int_{0}^{2\pi} d\theta / \rho(\theta)$ and $\tilde{\Omega}= \Omega/J_R$. 
Here energies are rescaled in units of $\hbar^2/(mR^2)$, $\tilde{\mu}$ is the rescaled chemical potential, $\tilde{g}$ is the interaction strength, $\tilde{V}(\theta) =  \tilde{V}_0\sum_{j=1}^n \exp[-2(\theta-\theta_j)^2/\sigma^2]$ is the sum of Gaussian barriers centered at $\theta_j = 2\pi j/n$, and $\theta$ is the azimuthal angle along the ring.
The free parameters $\tilde{g}$, $\sigma$ and $\tilde{V}_0$ are chosen in order to match the experimental conditions:  
$\sigma/\xi = 1.2$, $\tilde{V}_0/\tilde{\mu}_0 = 1.4$ and $\xi/R = 0.056$ (with $R = 12$ $\mu$m being approximately the inner radius of the experimental system), where $\mu_0$ is the chemical potential obtained in the homogeneous case (without barriers) and for $w=\Omega=0$.
For a given number of barriers, the solution of Eq.~(\ref{GPEstationary}) is obtained by imaginary time evolution.

{\it 3D GPE.}
In order to better capture the experimental procedure and the dynamics of the system, in 3D we solve numerically the time-dependent GPE for static barriers, 
\begin{equation}
    i\hbar \frac{\partial  \psi (\mathbf{r},t)}{ \partial t}=-\frac{\hbar ^2}{2M} \nabla ^2 \psi (\mathbf{r},t)+V(\vect{r}) \psi(\mathbf{r},t) +g |\psi(\mathbf{r},t)| ^2 \psi(\mathbf{r},t),
    \label{eq.GP.methods}
\end{equation}
with $\psi(\mathbf{r},t)$ being the condensate order parameter, $M$ the molecule mass, $g=4 \pi \hbar^2 a/M$ the interaction strength, $a=1010 \, a_0$ the s-wave scattering length and $a_0$ the Bohr radius.
The external trapping potential is $V(\vect{r}) = V_{\mathrm{harm}}(\vect{r})+V_{\rm ring}(\vect{r})+V_{\rm barr}(\vect{r})$. 
Here, $V_{\mathrm{harm}}(\vect{r})= M ( \omega_\perp^2 r^2 + \omega_z^2 z^2)/2$ is an harmonic confinement with $\{\omega_\perp , \, \omega_z \}= 2\pi \, \times \{2.5 \, , \, 396\}$ Hz.
The hard-wall potential creating the ring confinement in the $x$-$y$ plane is given by
\begin{equation}
\displaystyle
V_{\rm ring}(\vect{r})=V_r \left[\tanh\left(\frac{r-R_{\mathrm out}}{d}\right)+1\right]+V_r \left[\tanh\left(\frac{R_{\rm in}-r}{d}\right)+1\right].
\label{Vbound}
\end{equation}
with $R_{\mathrm in}=10.09 \, \mu \text{m}$ and $R_{\mathrm out}=21.82 \, \mu \text{m}$ 
being the inner and outer radius, respectively. 
The parameter $d=1.1 \, \mu \text{m}$ characterizes the stiffness of the hard walls, fixed such that the numerical density profiles match the in-situ experimental ones. 
We take $V_r$ larger than the chemical potential $\mu$ such that the density goes to zero at the boundary. 
The $n$ barriers are modelled as identical Gaussian peaks of trapping potential
\begin{equation}
\displaystyle
V_{\rm barr}=V_0 \sum _{i=1} ^{n/2} \exp{\left[-2(x\cos(i 2 \pi/n)+y \sin(i 2\pi/n))^2/{\sigma}^2\right]}.
\label{Vbarr}
\end{equation}
with constant width $\sigma=0.8 \, \mu $m.
We first find the system ground state by solving the GPE by imaginary time evolution and in the presence of $n$ barriers. 
We then instantaneously imprint a current of winding $w_0$ by multiplying the ground state wavefunction by the phase factor $\exp(-i 2 \pi w_0 \theta)$, where $\theta$ is the azimuthal angle. 
We finally study the system dynamics by solving the time-dependent GPE.  
For a particle number $N=6.8\times 10^3$ (corresponding to the experimental condensate number), we  obtain $\mu=1.09$ kHz leading to a value of the healing length  $\xi=0.59 \, \mu \text{m}$.
Equation~\ref{eq.GP.methods} is solved numerically by the Fourier split-step method on a Cartesian grid of $\{ N_x , N_y , N_z \} = \{ 256 , 256 , 80\}$ points dividing  a grid size of length $-34.846 \, \mu \text{m} \le r \le 34.846 \,  \mu \text{m}$ and $-11.0 \, \mu \text{m}  \le z \le 11.0 \, \mu \text{m}$ in the radial plane and axial direction, respectively. 
The time step is set to $\Delta t = 1 \times 10^{-5}\, \omega_\perp^{-1}$.

\begin{figure}[t!]
\includegraphics[width=1\columnwidth]{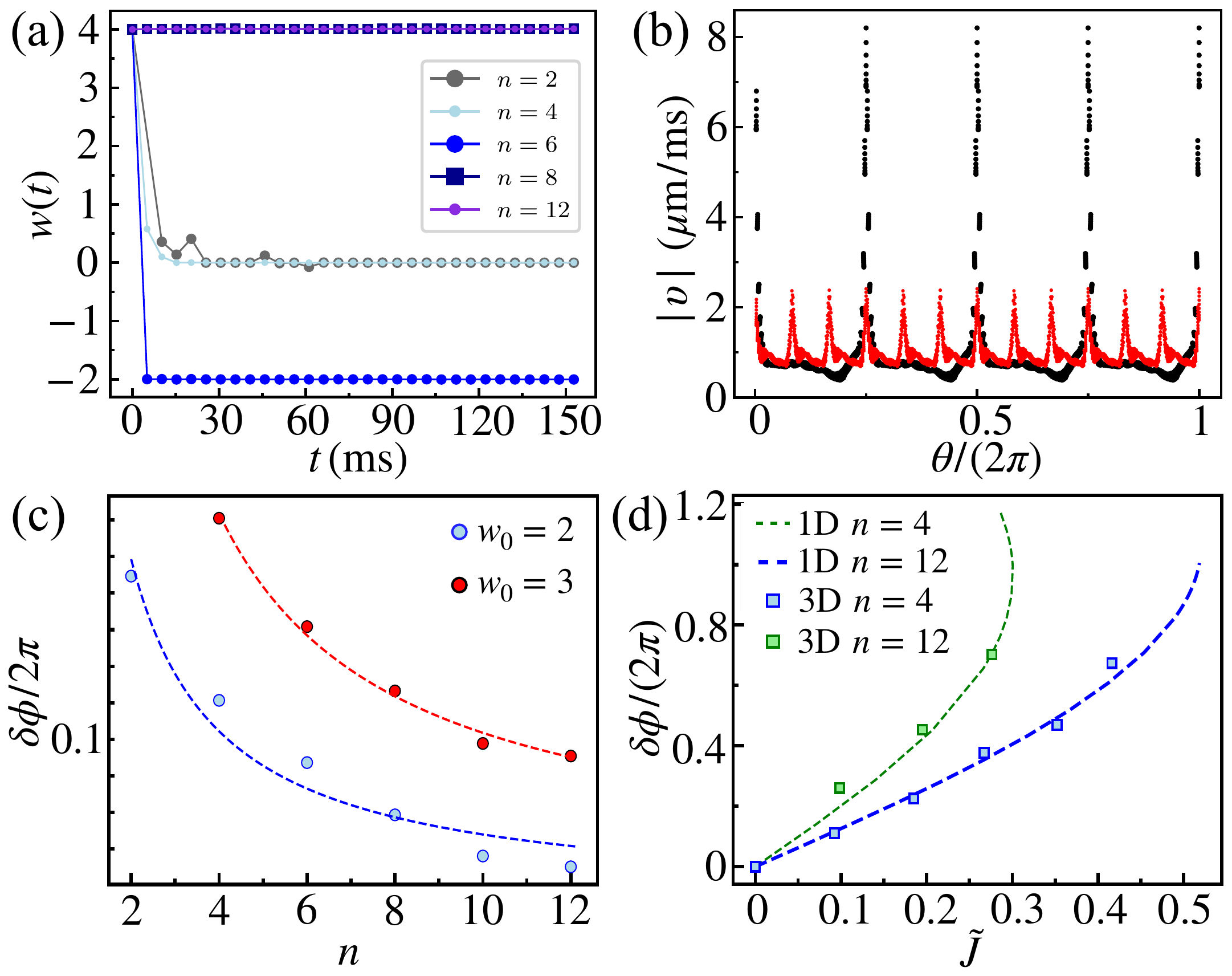}
\caption{{Results of the 3D GPE numerical simulations.
(a) Winding number as a function of time for fixed  $w_0=4$ and at different values of $n$ (see legend). 
For the parameters considered in these simulations $n_c(w_0)=8$.
(b) Absolute value of the superfluid velocity extracted at the mean radius $R=15$ $\mu$m for $z=0$, as a function of the azimuthal angle $\theta$.
Black (red) lines as obtained for $n=4$ ($n=12$), an initial winding number $w_0=2$ and at time $t=5\, $ ms. (c) Time averaged phase gain across each junctions as a function of the number of barriers $n$ and for different values of $w_0$. 
The dashed line is the fit function that goes as $1/n$.
(d) Time-averaged phase-current values extracted from the 3D time-dependent GPE simulations for $n=4$ (green squares) and $n=12$ (blue). 
The dashed lines are the corresponding phase-current curves, obtained with 1D simulations [as in the inset Fig.~\ref{Figure2}(a)].}
}
\label{Figure5}
\end{figure}

{\bf 3D simulations results.}
We characterise the condensate dynamics by studying the winding number, $w(t)$ calculated at at $z=0$ and averaged over closed circle paths ranging from the inner to the outer radius. 

\begin{figure}[t!]
\includegraphics[width=0.98\columnwidth]{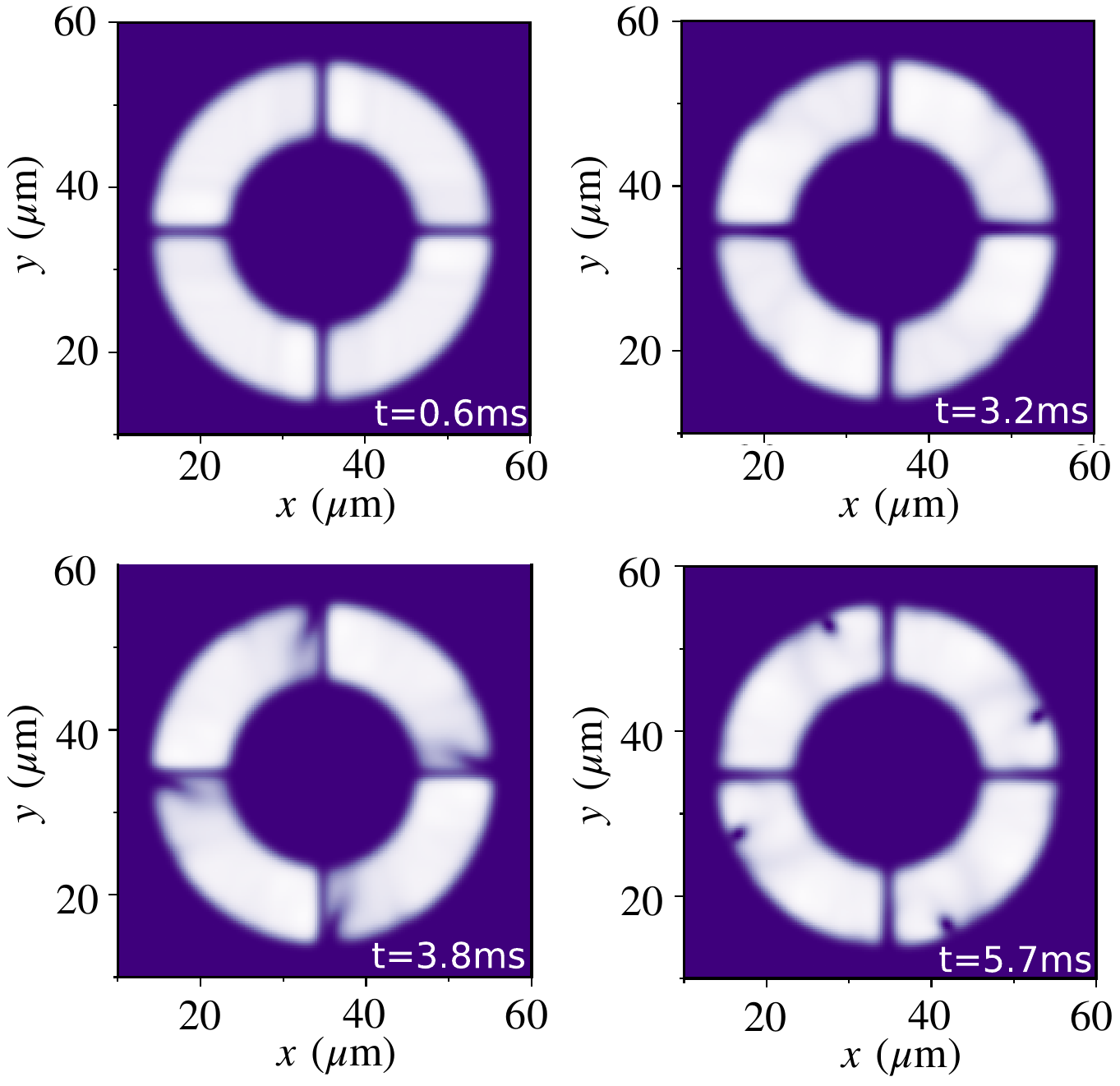}
\caption{Density profiles obtained for an unstable configuration of the system ($w_0=4$, $n=4$ and barrier height $V_0/\mu=1.4$) from 3D GPE numerical simulations. Each panel represents the superfluid density integrated along the $z$-direction.
The figure clearly shows the simultaneous nucleation of $n$ vortices, each vortex being emitted from the junction edge.
}
\label{Figure6}
\end{figure}

{\it Stable configuration.}
As discussed in the main text, for a fixed initial circulation $w_0$  we find the transition from unstable (decaying $w(t)$) to stable current (time-independent evolution of the winding) when the number of barriers $n$ exceeds a critical value $n_c(w_0)$,  see Fig.~\ref{Figure5}(a). 
In the stable configuration, 3D simulations reproduce the typical finding of the 1D case, namely that both the maximum of the superfluid speed and the phase gain at each junction decrease with $n$ as shown in Fig.~\ref{Figure5}(b)-(c).
In particular, figure~\ref{Figure5}(b) shows the absolute value of the superfluid velocity, $\vert  \vect{\upsilon}(\vect{r},t) \vert$, computed at the mean radius and for $z=0$, as a function of the azimuthal angle $\theta$. 
Figure ~\ref{Figure5}(c) instead shows the time-averaged phase gain across each junction, $\delta\phi$, as a function of $n$ (symbols), together with a $1/n$ fit (dashed lines). 
Finally, in Fig.~\ref{Figure5}(d), we plot the time-averaged $\delta \phi$ as a function of the time-averaged current (symbols).
The dotted lines are the 1D current-phase relation obtained for the same number of junctions and rescaled to be consistent with the data.
We see that the 3D results are consistent with the trend of an increasing critical current with $n$ found in 1D.

{\it Unstable configuration.}
If $n<n_c(w_0)$, we find that both $w$ and the current decay in time via vortex emission.
In Fig.~\ref{Figure6} we show the numerical densities illustrating the microscopic mechanism of the vortex emission process. Vortices are emitted symmetrically from each barrier: 
they enter the ring from the central part, propagate along the transverse direction close to the barrier position until they enter the bulk and travel at the outer edge of the ring. 
Each vortex entering the bulk through the barrier causes a global decrease of the winding number by one. 
In particular, for the considered case of $n=4$, the winding at $t=5.7$ ms is equal to zero. 
We note that the detailed vortex emission process depends on the value of the barrier height considered.

\begin{figure}[t!]
\includegraphics[width=0.98\columnwidth]{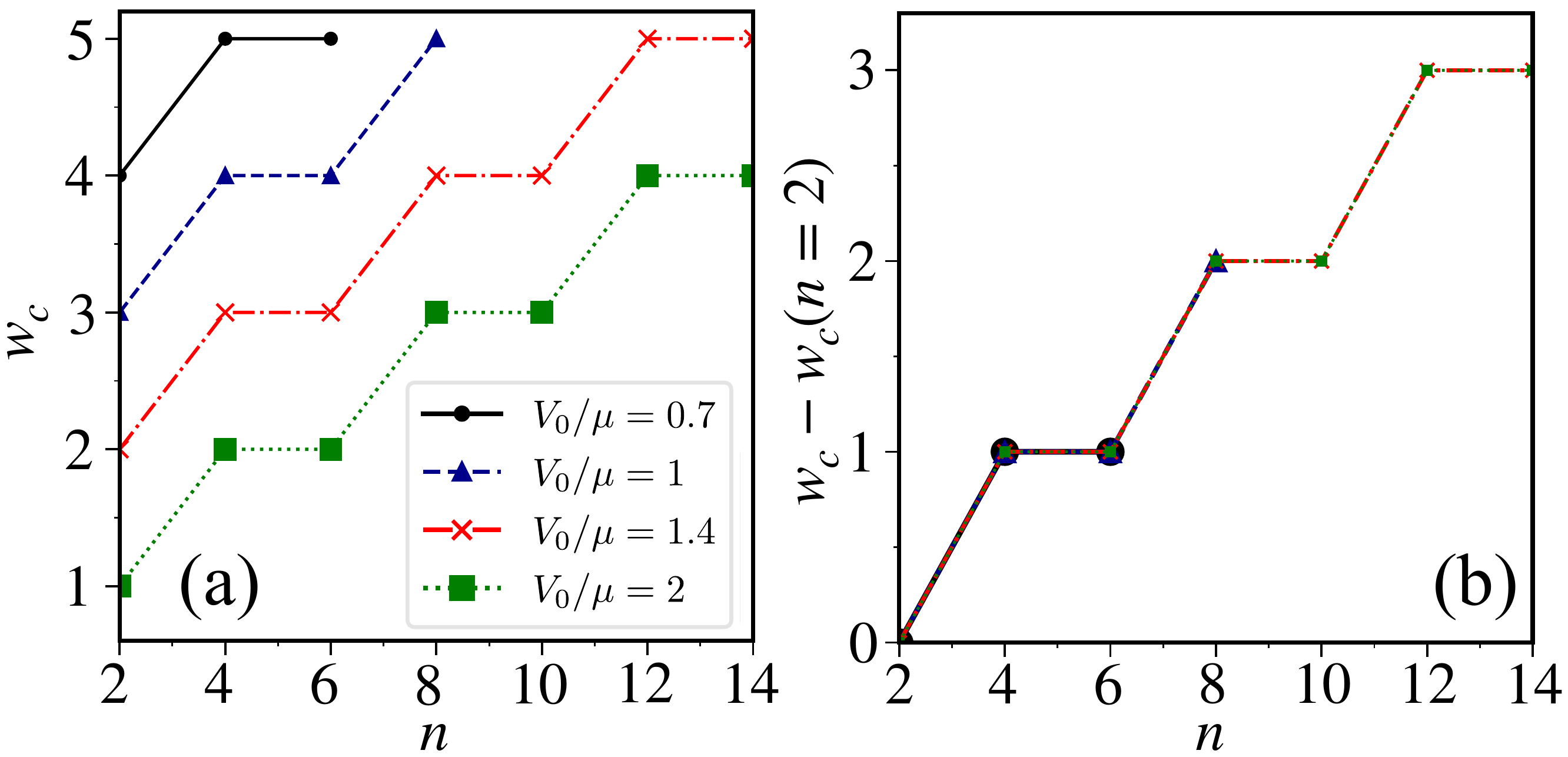}
\caption{Effect of the barrier heights on the critical circulation.
(a) Critical circulation $w_c(n)$ as a function of $n$ and for different values of the barrier height $V_0/\mu$.
(b) By substracting from $w_c(n)$ the critical value for $n=2$, the various curves collapse onto a single curve and we obtain a curve independent of $V_0/\mu$.
}
\label{Figure7}
\end{figure}

{\it Critical circulation.}
The stability phase diagram can be either characterized by the  critical number of barriers $n_c(w_0)$ for fixed $w_0$, as discussed above, or by the critical circulation $w_c(n)$.
For a given $n$, the critical circulation is the largest value of $w_0$ for which we find a stable dynamics.  
In Fig.~\ref{Figure7} we plot $w_c(n)$ as a function of $n$ and for different values of $V_0/\mu$.
Interestingly, all the curves are parallel to each other and thus collapse on the same curve upon a rescaling by $w_c$. 
In particular, by increasing $V_0/\mu$ the value of the critical circulation is shifted downwards. 
This is in agreement with the results in Ref.~\cite{xhani2023decay} for a single defect, which found that $w_c$ becomes smaller at larger values of $V_0/\mu$.
The dashed white line in Fig.~\ref{Figure4}(b) of the main text is obtained by using $V_0/\mu$ as fitting parameter. 
We obtain that experimental data are well reproduced by $V_0/\mu = 1.80 \pm 0.05$. 
This value is larger than the experimental estimation of $V_0/\mu = 1.3$, even after taking into account a decrease by 15\% of the chemical potential due to particle losses. 
The reason behind this discrepancy is investigated in the following paragraph.

\begin{figure}[b!]
\includegraphics[width=0.98\columnwidth]{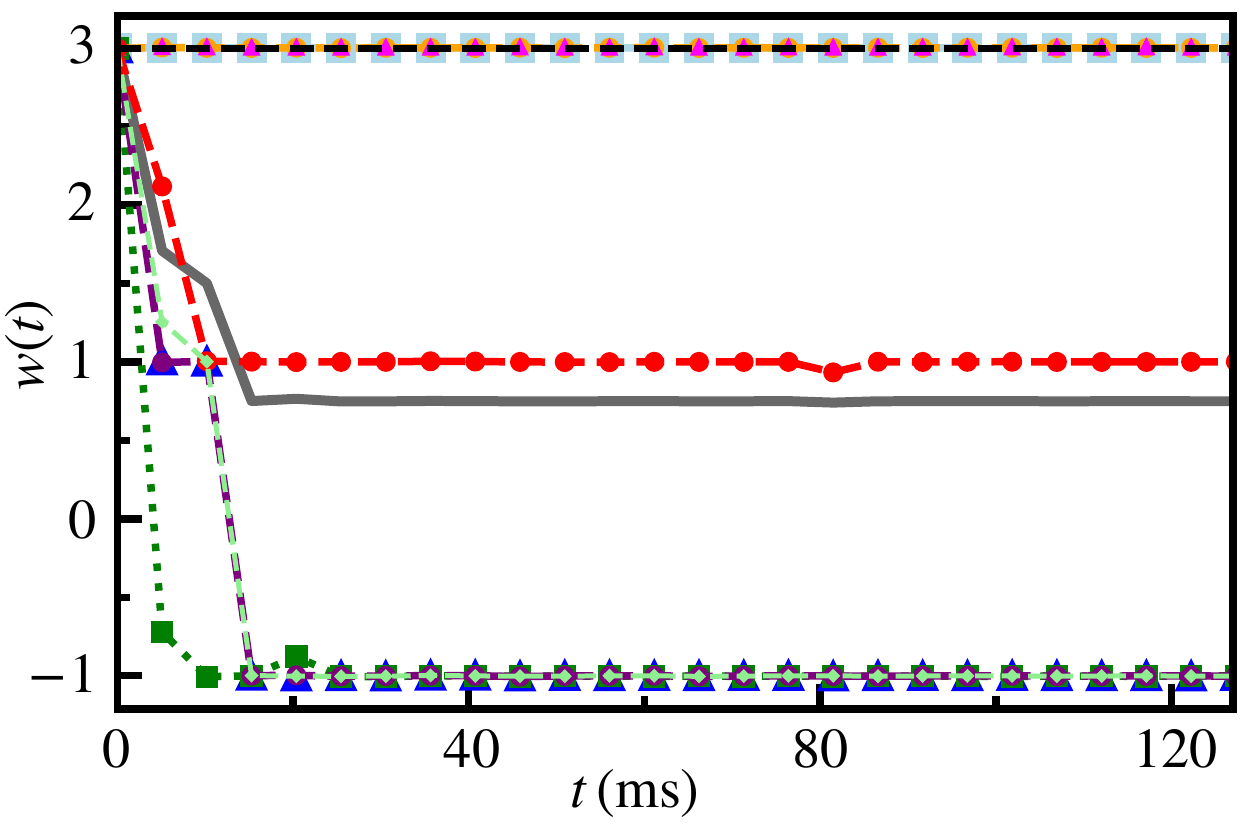}
\caption{{Winding number as a function of time for eight different simulations run with different barrier configurations (symbols with guides to the eye) and their average value (grey line). 
In each run, each barrier height and width is chosen randomly in the range $V_0/\mu_0=[1.2, 1.6]$
and $\sigma = [0.68, 0.92]$ $\mu$m, respectively. 
The black dashed line show the stable winding number in the absence of noise, i.e. for identical barriers of height $V_0/\mu_0 = 1.4$ and width $\sigma=0.8$ $\mu$m. 
The data are produced from 3D numerical simulations in the case of $n=4$ barriers and at fixed $w_0=3$.}
} \label{Figure8}
\end{figure}

\begin{figure*}[t!]
\includegraphics[width=1\textwidth]{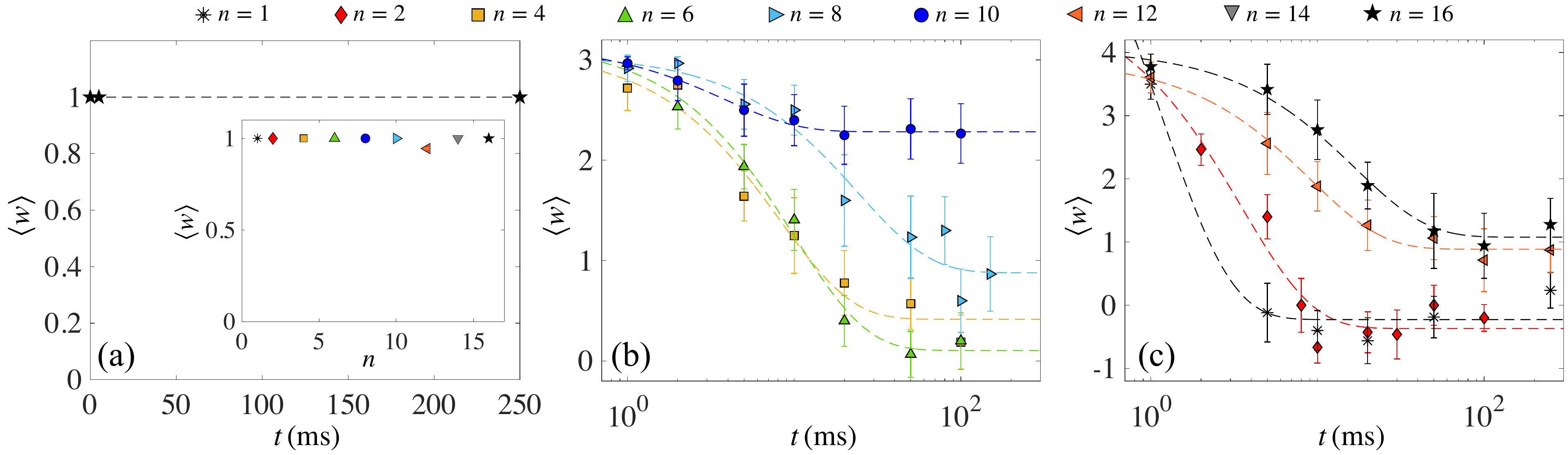}
\caption{Statistically-averaged winding number as a function of time and for different $n$ (symbols). 
The different panels refer to different values of $w_0$: $w_0=1$ (a), $w_0=3$ (b) and $w_0=4$ (c).
Dashed lines represent the exponential fit of each dataset, using the same fitting function as in Fig.~\ref{Figure4}(a).
The inset of panel (a) reports the averaged winding number at $t=250$ ms as a function of $n$ for $w_0=1$.
}
\label{Figure9}
\end{figure*}

\begin{figure}[t!]
\includegraphics[width=1\columnwidth]{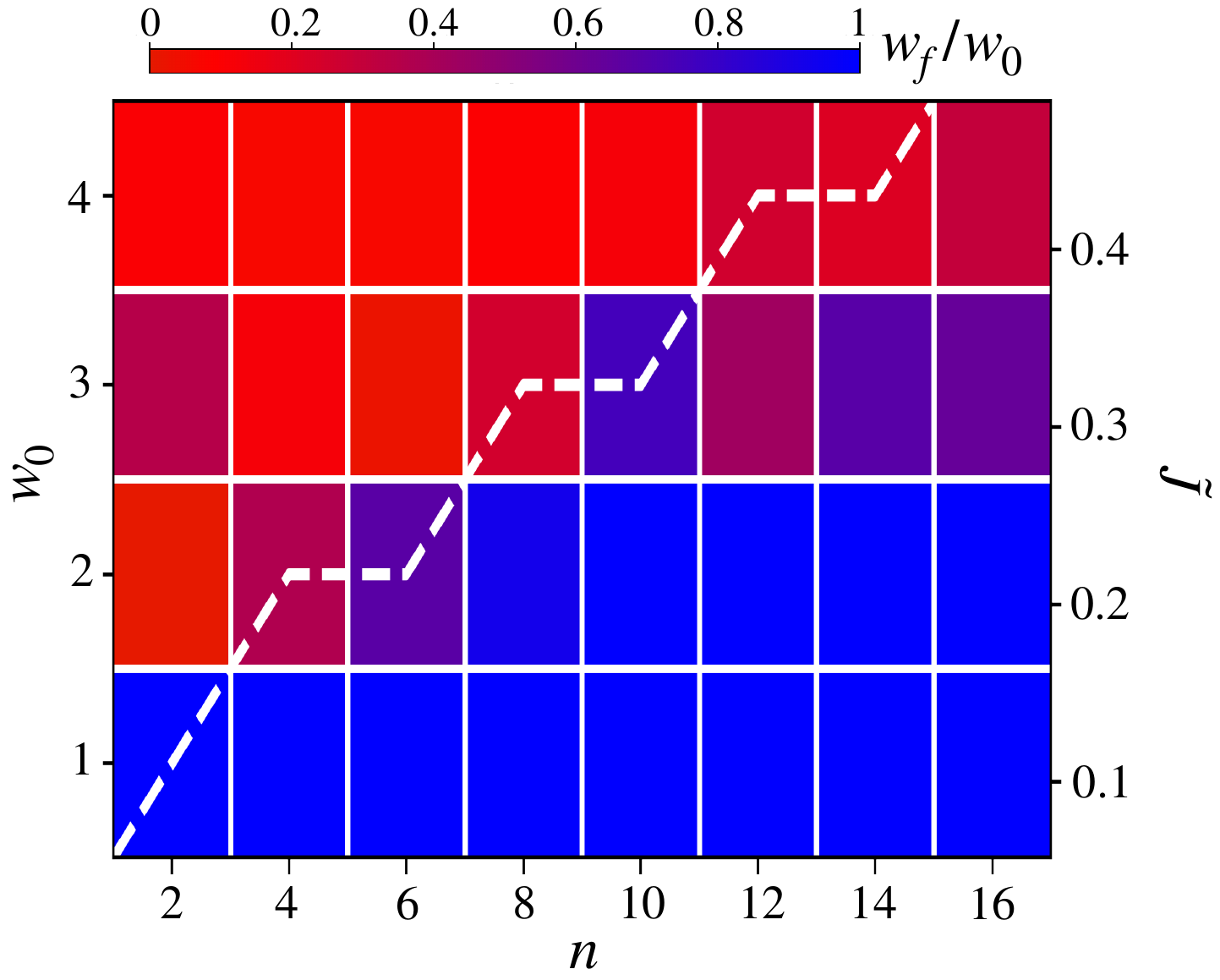}
\caption{Measured ratio $w_f/w_0$ (colormap) as a function of $w_0$ and $n$.
The white dashed line is the same as in Fig. ~\ref{Figure4}(b).
}
\label{Figure10}
\end{figure}

{\it Effect of non-identical barriers on the current's stability.} 
Due to the finite resolution of the DMD-created optical potentials, the experimental barriers creating the JJN are not identical, but their height and size is distributed around the mean values mentioned in the main.
To check whether non-identical barriers could affect the current stability in the JJN, we study this case with 3D numerical simulations.
In particular, in Fig.~\ref{Figure8} we report the time evolution of the winding for $n=4$ barriers and $w_0=3$, which is stable for identical barriers, when each barrier height and width are randomly selected from Gaussian distributions of mean values $V_0/\mu_0=1.4$ and $\sigma=0.8 \, \mu$m and standard deviation $\Delta V_0/\mu_0=0.2$ and $\Delta \sigma=0.12$ $\mu$m, respectively. The mean and standard deviation values correspond to the ones measured from the experimental characterization of the barriers.
In the figure, we show the winding number as a function of time for 8 different runs (symbols), corresponding to different configurations of the barriers.
We also plot the statistical mean value (solid line).
These simulations reproduce qualitatively the experimental findings: for some barrier configurations, the winding number remains constant in time, for some others it decays, eventually also reaching negative values.
Correspondingly, also the average winding number (solid line) decays in time. 
To summarize, having non-identical barriers is observed to reduce the stability of currents in the JJN for the same $w_0$, explaining the discrepancy in $V_0/\mu$ found between the experimental and numerical phase diagram.
Finally, we have also
checked numerically -- by solving the collisionless Zaremba-Nikuni-Griffin model \cite{ZNGBOOK,XhaniPRR} for an experimentally estimated condensed fraction of $80\%$ -- that finite-temperature dissipation does not affect the critical winding number and it affects only slightly the decay time.

{\bf Experimental methods.}
We discuss here additional details of the methods
employed to obtain the experimental data presented in the main text.

{\it Characterization of the tunneling barriers.}
Due to the finite resolution of the DMD-projecting setup, the barriers of experimental JJNs are not identical. We characterize the properties of each barrier in the different configurations at various $n$ by acquiring an image of the DMD-created light profile by means of a secondary camera, and calibrating the optical potential via the equation of state of a BEC in a well characterized 3D harmonic trap~\cite{KwonSCIENCE2020}.
Then, we extract the height and $1/e^2$-width by fitting the radially-averaged profile of each barrier with a Gaussian.
From this set of data, we extract the mean values and standard deviation of barrier height $V_0 \simeq (1.3 \pm 0.2)\,\mu$ and width $\sigma = (1.2 \pm 0.2)\,\xi$.
Error bars denote the standard deviation of the parameters over the set of barriers.
Even though the barriers are not strictly identical, the obtained results show that it is possible to create similar barriers with 
fluctuations on $V_0$ and $\sigma$ that are only a fraction of the chemical potential and healing length, respectively.

\begin{figure}[t!]
\includegraphics[width=0.8\columnwidth]{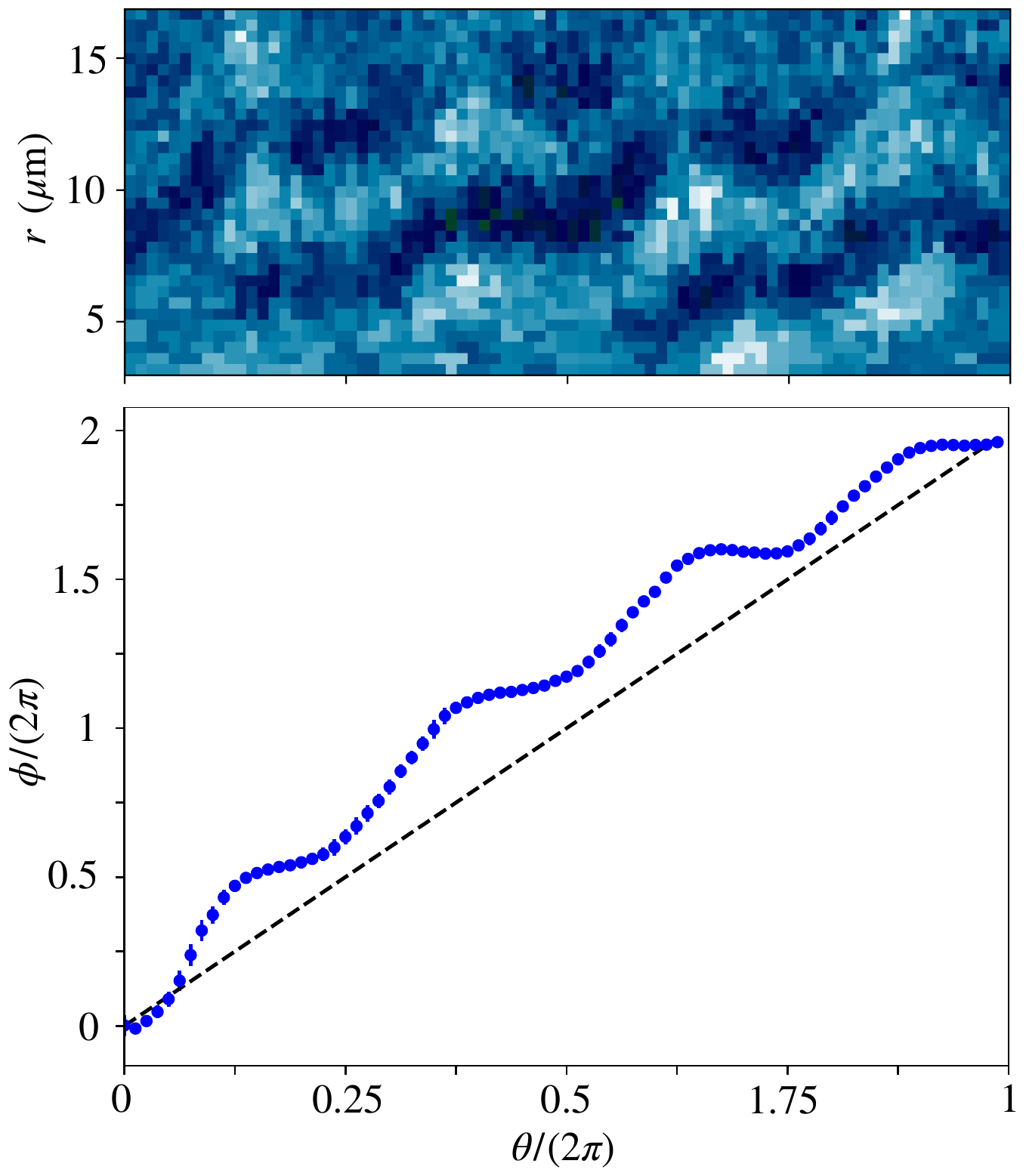}
\caption{Extraction of the azimuthal profile of the JJN phase from the experimental interferograms. 
The upper panel shows the interferogram image unwrapped into polar coordinates.
A sinusoidal fit of each azimuthal slice is performed by using the function $A \cos (\alpha r + \phi)$. 
To enhance the fringe contrast, for each value of $\theta$ we consider a radial slice of the polar interferogram 
averaged over $\Delta\theta = 0.47 \,$rad.
The image is obtained by averaging 5 similar experimental spiral patterns and unwrapping the resulting image.
The lower panel is the fitted azimuthal trend of $\phi$, showing a phase jump in correspondence of each barrier.
The unwrapped phase here is averaged over the profiles of all experimental images for $(w_0=2,\, n=4)$ where $w(t) = w_0$ at times $t>10$ ms.
}
\label{Figure11}
\end{figure}

{\it Experimental phase profile in the JJN.}
As already commented in the main text, the interferograms associated with stable realizations, namely in which $w = w_0$, show interference fringes with a clear polygonal structure (e.g. squared for $n = 4$), which are a manifestation of the phase jump at each Josephson junction. Thanks to the high resolution of the imaging setup, we can extract the local relative phase between the ring and the reference central disc, $\phi$, as a function of the azimuthal angle $\theta$, as reported in Fig.~\ref{Figure11}. The interferograms in polar coordinates [(a)] display a characteristic step-like shape of the fringes, closely resembling the predicted behavior of the JJN phase by the analytical model and from the numerical simulations (see the inset of Fig.~\ref{Figure2}(a) for comparison). We then quantitatively extract the value of $\phi(\theta)$ as the phase shift in the sinusoidal fit of a slice of the polar interferogram at constant $\theta$.
As shown in Fig.~\ref{Figure11} (b), the $\phi (\theta)$ trend clearly deviates from the linear behavior expected in a clean ring~\cite{DelPacePRX2022}, but it rather exhibits a number of jumps in correspondence of the barriers in the JJN.

{\it Experimental stability phase diagram.}
In Fig.~\ref{Figure9} we provide additional experimental data regarding the statistically-averaged winding number as a function of time and for different $n$ and $w_0$.
In the case $w_0=1$, $\langle w \rangle$ is found to be constant in time up to 250 ms for any $n$.
In particular, in Fig.~\ref{Figure9}(a) we plot the case $n=16$, averaged over about 20 realizations.
The inset of Fig.~\ref{Figure9}(a) shows $\langle w \rangle$ at time $t= 250$ ms and for $n$ ranging from 1 to 16. 
Only in the case $n=12$, we found a single experimental realization (out of 18 independent runs) with $w = 0$. 
In Fig.~\ref{Figure9}(b) and (c) we plot the cases $w_0=3$ and $w_0=4$ [the case $w_0=2$ is shown in Fig. \ref{Figure4}(a)]. 

In Fig.~\ref{Figure10} we report the analogue of the stability phase diagram of \ref{Figure4}(a) here plotting $w_f/ w_0$, where $w_f$ is the average circulation at long time, as obtained from a fit (see main text).
This shows the experimental deterministic realization of stable circulation states for $w=1$ and $w=2$ in a toroidal trap with up to $n=16$ junctions. 
\smallskip

\begin{figure}[t!]
\includegraphics[width=1\columnwidth]{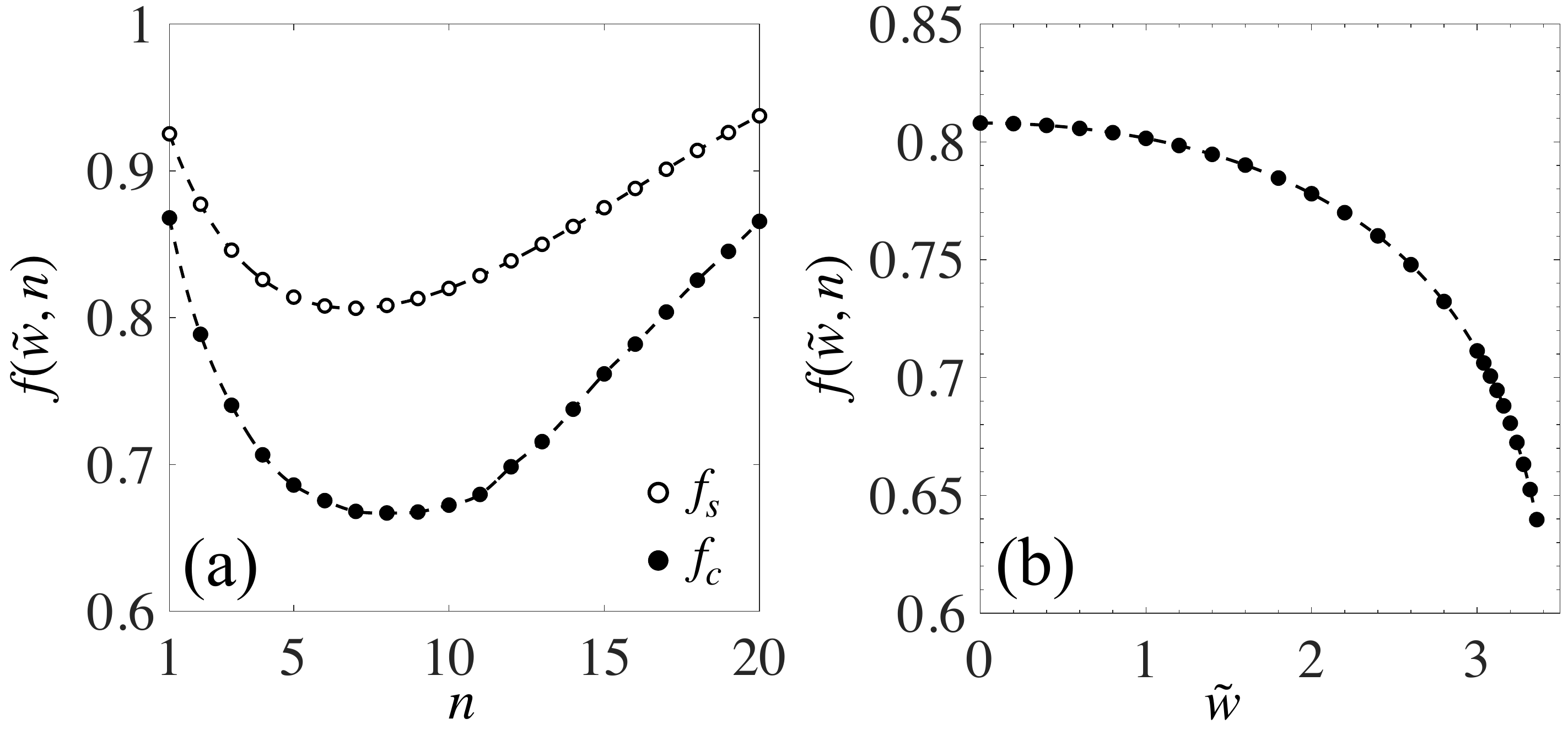}
\caption{Function $f(\tilde{w},n)$ calculated for stationary states of the 1D GPE.
Panel (a) plots $f(\tilde{w},n)$ as a function of $n$ for two interesting cases: $f_c(n)=f(\tilde{w}=\tilde{w}_c,n)$ (dots dots) and $\lim_{w=0, \Omega\to 0} f(\tilde{w},n) =f_s$ (circles), corresponding to Leggett's superfluid fraction, Eq.~(\ref{fs1Dcell}). 
Panel (b) shows $f(\tilde{w},n)$ as a function of $\tilde{w}$ and for $n=6$ (dots).
In both panels, lines are guides to the eye.
}
\label{Figure12}
\end{figure}

{\bf Superfluid fraction and the $f(\tilde{w},n)$ function.}
The superfluid fraction for neutral atoms in a ring trap rotating at an angular velocity $\Omega$ can be defined as~\cite{LeggettPRL1970, LeggettJSP1998}
\begin{equation}\label{fs_Leggett}
    f_s = 1-\lim_{\Omega \xrightarrow{} 0} \frac{L}{I_{\rm cl} \Omega}
\end{equation}
where $L$ is the expectation value of the angular momentum and  $I_{\rm cl}$ is the classical moment of inertia. 
In 1D, we have
\be \label{fs1Dcell}
f_s = \frac{1}{\frac{1}{d^2} \int_{\rm cell} \frac{d\theta}{\rho(\theta)}},
\ee
where $\rho(\theta)$ is normalized to one over the unit cell of azimuthal size $d$.
Equation~(\ref{fs1Dcell}) is derived by noticing that the two bounds in Eq.~(\ref{fs_bounds}) coincide in 1D.
In our case, restricting to the unit cell as in Ref.~\cite{LeggettPRL1970, LeggettJSP1998} is not necessary and Eq.~(\ref{fs1Dcell}) is unchanged if we write $f_s = \frac{1}{(2\pi)^2} [ \int \frac{d\theta}{\rho(\theta)}]^{-1}$ with $\rho(\theta)$ normalized to one over the full circle, even in the presence of $n$ junctions.
In particular, we have $f_s = \lim_{w=0,~ \Omega \to 0} f(\tilde{w},n)$, where $f(\tilde{w},n)$ is related to the current according to Eq.~(\ref{Eq.J}).
In Fig.~\ref{Figure12}(a) we plot $f_s$ (circles) and $f_c$ [corresponding to $f(\tilde{w}=\tilde{w}_c,n)$, dots] as a function of $n$.
Both functions decrease with $n$ until the barriers start to overlap.
In Fig.~\ref{Figure12}(b) we plot $f(\tilde{w},n)$ as a function of $\tilde{w}$ for $n=6$.

To compare numerical and experimental data in Fig.~\ref{Figure4}(c), we have taken into account the finite spatial resolution of the imaging system, characterized by a Point Spread Function (PSF) of full-witdh-half-maximum FWHM = $0.83 \, \mu$m \cite{KwonSCIENCE2020}. 
To estimate the theoretical curves of Fig.~\ref{Figure4}(c), we first integrate the 3D numerical densities along the z direction,  
Then, we account for the finite experimental resolution by convolving the integrated numerical densities with a two-dimensional Gaussian with a FWHM matching the experimental PSF.
This procedure leads to a decrease in the resolution of the density modulation, which causes the estimated superfluid fraction to increase and yields results in good agreement with experimentally extracted values [see Fig.~\ref{Figure4}(c)].

\end{document}